\documentclass[aps,pra,showkeys,twocolumn,superscriptaddress]{revtex4-2}
\usepackage{array}
\usepackage{booktabs}
\usepackage{tabu}
\usepackage{dcolumn}
\usepackage{amsmath}
\usepackage{amsfonts}
\usepackage{float}
\usepackage{amssymb}
\usepackage{graphicx,color}
\usepackage[colorlinks={true}]{hyperref}
\hypersetup{colorlinks=true,linkcolor=red,citecolor=blue,urlcolor=blue}
\usepackage{graphicx}
\usepackage{subfigure}
\usepackage{graphicx}
\usepackage{dcolumn}
\usepackage{bm}
\usepackage{pstricks}
\usepackage{braket}
\usepackage{orcidlink}

\def\be{\begin{equation}}
  \def\ee{\end{equation}}
\def\bea{\begin{eqnarray}}
\def\eea{\end{eqnarray}}
\def\f{\frac}
\def\n{\nonumber}
\def\l{\label}
\def\p{\phi}
\def\o{\over}
\def\R{\rho}
\def\pa{\partial}
\def\om{\omega}
\def\na{\nabla}
\def\P{\Phi}
\begin{document}

\title{Extracting Work From Two Gravitational Cat States}

\author{Maryam Hadipour \orcidlink{0000-0002-6573-9960}}
\affiliation{Faculty of Physics, Urmia University of Technology, Urmia, Iran}
\author{Soroush Haseli \orcidlink{0000-0003-1031-4815}}\email{soroush.haseli@uut.ac.ir}
\affiliation{Faculty of Physics, Urmia University of Technology, Urmia, Iran}

\date{\today}
\def\be{\begin{equation}}
  \def\ee{\end{equation}}
\def\bea{\begin{eqnarray}}
\def\eea{\end{eqnarray}}
\def\f{\frac}
\def\n{\nonumber}
\def\l{\label}
\def\p{\phi}
\def\o{\over}
\def\R{\rho}
\def\pa{\partial}
\def\om{\omega}
\def\na{\nabla}
\def\P{$\Phi$}

\begin{abstract}
This work examines how a thermal environment affects the work that can be extracted from gravitational cat states. The purpose of this work is to provide an in-depth discussion of the effects of temperature and gravitational interaction between states with masses $m$ on work extraction. The results show that the increase in temperature and the interaction between states decrease the amount of work that can be extracted from gravitational cat states.
\end{abstract}

\keywords{Quantum gravity; Ergotropy system; Cat states}

\maketitle

\section{Introduction}	
During the last few decades, gravity and quantum field theory have increasingly focused on the quantization of gravitational interactions. The main focus of these efforts have been to quantitatively quantize the gravitational field or space-time geometry \cite{1,2,3,4}. Due to technological advances, phenomenological approaches to quantum gravity have emerged in recent years that have led to experimental bounds on estimations of Planck Scale \cite{5,6}. Recent investigations have taken into account underground and laboratory-based measurements \cite{7,8,9,10} as well as constraints from observations of astrophysical events \cite{11,12}. Recently, fundamental question has been raised in this case, and that is, does the gravitational interaction really need to be quantized or not \cite{13}? Furthermore, do we have a model-independent experimental protocol that can confirm gravity is an interaction of quantum mechanical nature \cite{14,15}? Considering that the origin of quantum entanglement phenomenon  is the interaction between two systems and considering that the gravitational interaction leads to the creation of entanglement between two states, it can be said that gravity has a quantum nature. So, the creation of quantum entanglement induced by gravitational interactions is an adequate  evidence of quantum gravity \cite{16,17,18}.  In this work, we do not intend to talk about the quantum correlations created by the gravitational interaction, however we want to study how the thermal bath affects the work that can be extracted from gravitational cat states.

A great deal of attention has been paid in recent studies to the extraction of work from quantum systems. \cite{19,20,21,22}. In this work, we intend to study the effects of temperature and gravitational interaction on the work that can be extracted from gravitational cat states in the presence of a thermal bath. It will be shown that the work that can be extracted from the gravitational cat states decreases with the increase in the temperature of the heat bath. It will also be shown that the amount of work that can be extracted decreases with the increase of gravitational interaction between states.
\section{Gravitational cat state in thermal environment}
\begin{figure}[H]
\centering
    \includegraphics[width =0.65 \linewidth]{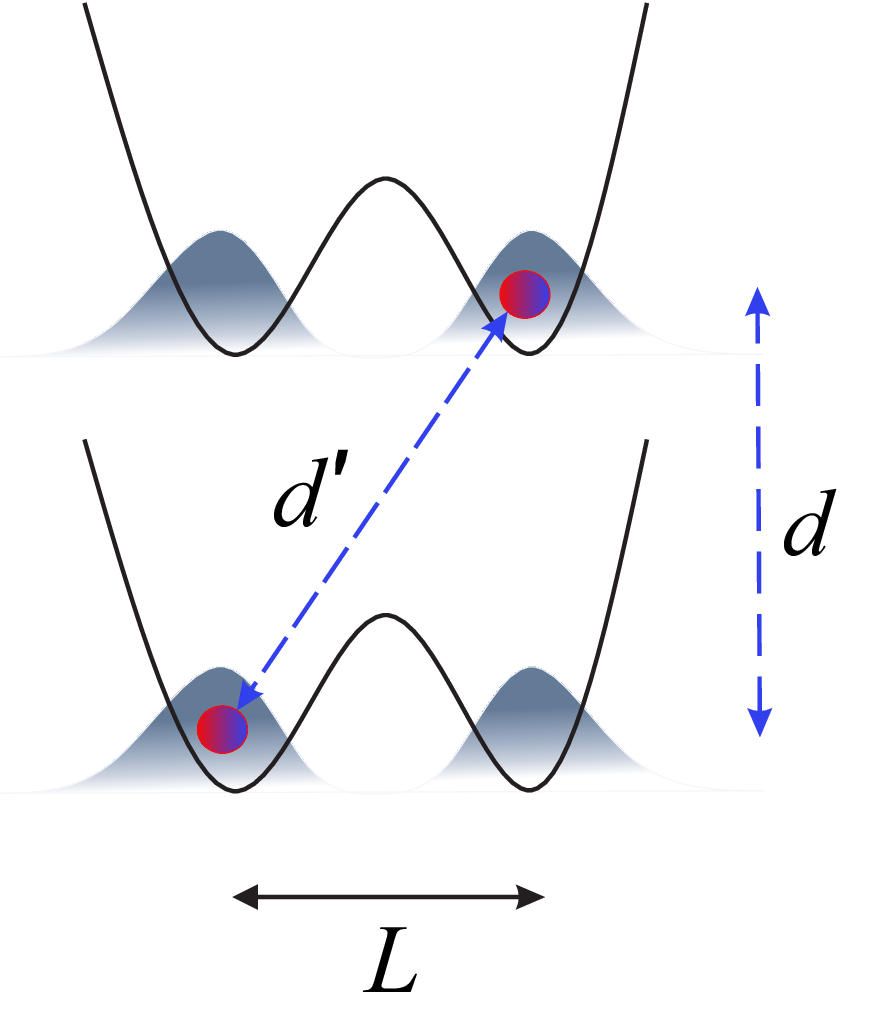}
    \caption{A visual representation of the considered set-up for gravitational cat states.}
    \label{Fig1}
  \end{figure}

Let us consider two particles of mass $m$, each of them in a one-dimensional double-well potential, with a local minimum at $x=\pm L/2$, this state is known as gravitational cat states. The potential is even and we can assign an eigenstate to each particle that is located in each of the minimum location. The assigned eigenstates $\vert \pm \rangle$, satisfy the $\hat{x} \vert \pm \rangle = \pm \frac{L}{2} \vert \pm \rangle$, where $\hat{x}$ is the position operator. Based on the Landau-Lifschitz approximation the mentioned eigenstates can be written as  $\vert \pm \rangle = 1/\sqrt{2}(\vert g \rangle \pm \vert e \rangle)$, where $\vert g \rangle$ and $\vert e \rangle$ are the ground and excited states with energy scale $-\omega/2$ and $\omega/2$, respectively. The schematic diagram of the model is shown in Fig.\ref{Fig1}. 
The Hamiltonian of this model can be written as \cite{23,had,Roj} 
\begin{equation}\label{Hamil}
\mathcal{H}=\frac{\omega}{2}(\sigma_z \otimes \mathit{I} + \mathit{I}\otimes \sigma_z)-\Omega (\sigma_x \otimes \sigma_x),
\end{equation}   
where $\sigma_x$, $\sigma_z$ are Pauli operators along $x$ and $z$ respectively. $\Omega$ quantifies the strength of the gravitational interaction between states with masses $m$. Gravitational interaction parameter $\Omega$ cen be obtained as $\Omega=\frac{\eta}{2}(\frac{1}{d}-\frac{1}{d^{\prime}})$, where $\eta=G m^2$ and $G$ is a universal gravitational constant. $d$ and $d^{\prime}$ are distance between the two masses when each of the is at the same and different relative minimum respectively.  From the Hamiltonian in Eq. \ref{Hamil}, eigenvalues can be obtained as 
\begin{equation}
\epsilon_1=-\Delta, \quad \epsilon_2=-\Omega,\quad \epsilon_3=\Omega, \quad \epsilon_4=\Delta
\end{equation} 
where, $\Delta=\sqrt{\Omega^{2}+ \omega^2}$. For each eigenvalue the associated eigenstate can be written as 
\begin{equation}
\begin{aligned}
\left|\epsilon_1\right\rangle & =\sin \left(\phi_{+}\right)|ee\rangle+\cos \left(\phi_{+}\right)|gg\rangle, \\
\left|\epsilon_2\right\rangle & =\quad \frac{1}{\sqrt{2}}(|ge\rangle+|eg\rangle), \\
\left|\epsilon_3\right\rangle & =\frac{1}{\sqrt{2}}(|ge\rangle-|eg\rangle),\\
\left|\epsilon_4\right\rangle & =\sin \left(\phi_{-}\right)|ee\rangle+\cos \left(\phi_{-}\right)|gg\rangle, 
\end{aligned}
\end{equation}
where $\phi_{\pm}=\arctan \left( \frac{\Omega}{\omega \pm \Delta} \right) $. To investigate the thermal effects, the system is considered to be in a thermal environment. In the thermal equilibrium the density matrix of quantum system can be described by density operator $\rho(T)=\exp(\frac{\mathcal{H}}{k_B T})/\mathcal{Z}$, where $k_B$ is the Boltzmann’s constant, $T$ is  absolute temperature and $\mathcal{Z}=tr (\exp(\frac{\mathcal{H}}{k_B T}))$ is the partition function of the system. Thermal state of considered system can be defined by following density matrix in computational basis 
\begin{equation}\label{den}
\rho=\left( \begin{array}{cccc}
\rho_{11} & 0 & 0 & \rho_{14} \\
0 & \rho_{22} & \rho_{23} & 0 \\
0 & \rho_{23} & \rho_{22} & 0 \\
\rho_{14} & 0 & 0 & \rho_{44}
\end{array}\right) ,
\end{equation}
the element of density matrix are given by
\begin{equation}
\begin{aligned}
& \rho_{11}=\frac{1}{Z}\left(\mathrm{e}^{-\beta \varepsilon_2} \sin ^2\left(\phi_{-}\right)+\mathrm{e}^{-\beta \varepsilon_3} \sin ^2\left(\phi_{+}\right)\right), \\
& \rho_{14}=\frac{1}{Z}\left(\frac{\mathrm{e}^{-\beta \varepsilon_2} \sin \left(2 \phi_{-}\right)+\mathrm{e}^{-\beta \varepsilon_3} \sin \left(2 \phi_{+}\right)}{2}\right) \text {, } \\
& \rho_{22}=\quad \frac{1}{Z}\left(\frac{\mathrm{e}^{-\beta \varepsilon_1}+\mathrm{e}^{-\beta \varepsilon_4}}{2}\right), \\
& \rho_{23}=\quad \frac{1}{Z}\left(\frac{\mathrm{e}^{-\beta \varepsilon_1}-\mathrm{e}^{-\beta \varepsilon_4}}{2}\right), \\
& \rho_{44}=\frac{1}{Z}\left(\mathrm{e}^{-\beta \varepsilon_2} \cos ^2\left(\phi_{-}\right)+\mathrm{e}^{-\beta \varepsilon_3} \cos ^2\left(\phi_{+}\right)\right), \\
&
\end{aligned}
\end{equation}
where $\beta=1/k_BT$.
\section{Work extraction from gravitational cat states in thermal bath} 
The extraction of work is one of the most important tasks in thermodynamics \cite{24}. The extraction of work from quantum systems has been the subject of many different studies \cite{24,25,26,27,28}. First, we start by reviewing the notion of ergotropy, which is the maximum amount of work that a unitary cyclic operations can extract from a quantum system. Starting from the assumption that the quantum system is thermally isolated from its surroundings and does not exchange heat with them. In addition, the process is cyclic, i.e., the system returns to its initial Hamiltonian at the end of the process. Almost any such process can be described by a unitary operation 
\begin{equation}
U(t)=\exp\left( -i \int_0^t ds \left[ \mathcal{H} + V(s) \right] \right),
\end{equation}
where $\mathcal{H}$ and $V(s)$ are the Hamiltonian and time dependent field that can be used to extract energy from the quantum system respectively. Due to the cyclical nature of the process, $V(s)$ disappears at the beginning and at the end of the process i.e. $V(0)=V(t)=0$. The work that can be extracted by such a process is given by
\begin{equation}
W=tr(\mathcal{H} \rho)- \min_U tr(\mathcal{H} U \rho U^{\dag}),
\end{equation}
where $\rho$ is density operator describing the state of the quantum system. $V(s)$ can be chosen properly to generate any unitary transformation $U$. So, the maximum value of the work that can be extracted from quantum system (ergotropy) is 
\begin{equation}
\mathcal{W}=tr(\mathcal{H} \rho)- \min_U tr(\mathcal{H} U \rho U^{\dag}),
\end{equation}
where the optimization is done over all possible unitary operations. It has been represented that there exist a unique state known as passive that maximized the above relation for each $\rho$ \cite{31,32,33}.  
\begin{equation}
\mathcal{W}=tr(\mathcal{H}\rho)-tr(\mathcal{H}\xi),
\end{equation}
where $\xi$ is passive state of $\rho$. Passive state has non-increasing population associates to its Hamiltonian and $\left[ \mathcal{H}, \eta \right] $. Using spectral decomposition the density matrix $\rho$ and Hamiltonian $\mathcal{H}$ of quantum systems can be written as
\begin{equation}
 \rho =\sum_n r_n \vert r_n \rangle \langle r_n \vert, \quad r_1 \geq r_2 \geq... \geq r_n 
\end{equation}
and 
\begin{equation}
\mathcal{H}=\sum_n \epsilon_n \vert \epsilon_n \rangle \langle \varepsilon_n \vert, \quad \varepsilon_1 \leq \epsilon_2 \leq ... \leq \epsilon_n,
\end{equation} 
where $\epsilon_n$ ($\vert \epsilon_n \rangle$ ) and $r_n$ ($\vert r_n \rangle$) are eigenvalues (eigenstate)of the $\mathcal{H}$ and $\rho$ respectively. So, the passive sate $\eta$ can be written as \cite{31} 
\begin{equation}
\eta=U \rho U^{\dag}= \sum_n r_n \vert \epsilon_n \rangle \langle \epsilon_n \vert.
\end{equation}
One can obtain the ergotropy as 
\begin{equation}
\mathcal{W}=\sum_{n,m} r_n \varepsilon_m \left( \vert \langle r_n \vert \varepsilon_m \rangle \vert^2 - \delta_{m,n} \right),
\end{equation}
where $\delta_{n,m}$ is the Kronecker delta function. Now by making use of the density matrix $\rho$ in Eq.\ref{den}, which describes the gravitational cat state in thermal bath and considering the Hamiltonian of the model $\mathcal{H}$ in Eq.\ref{Hamil}, the amount of the work that can be extracted from gravitational cat state in thermal bath can be obtained easily. Here, the effect of temperature and the gravitational interaction strength on the work extraction from  gravitational cat state will be studied. 
\begin{figure}[t]
\centering
    \includegraphics[width =0.8 \linewidth]{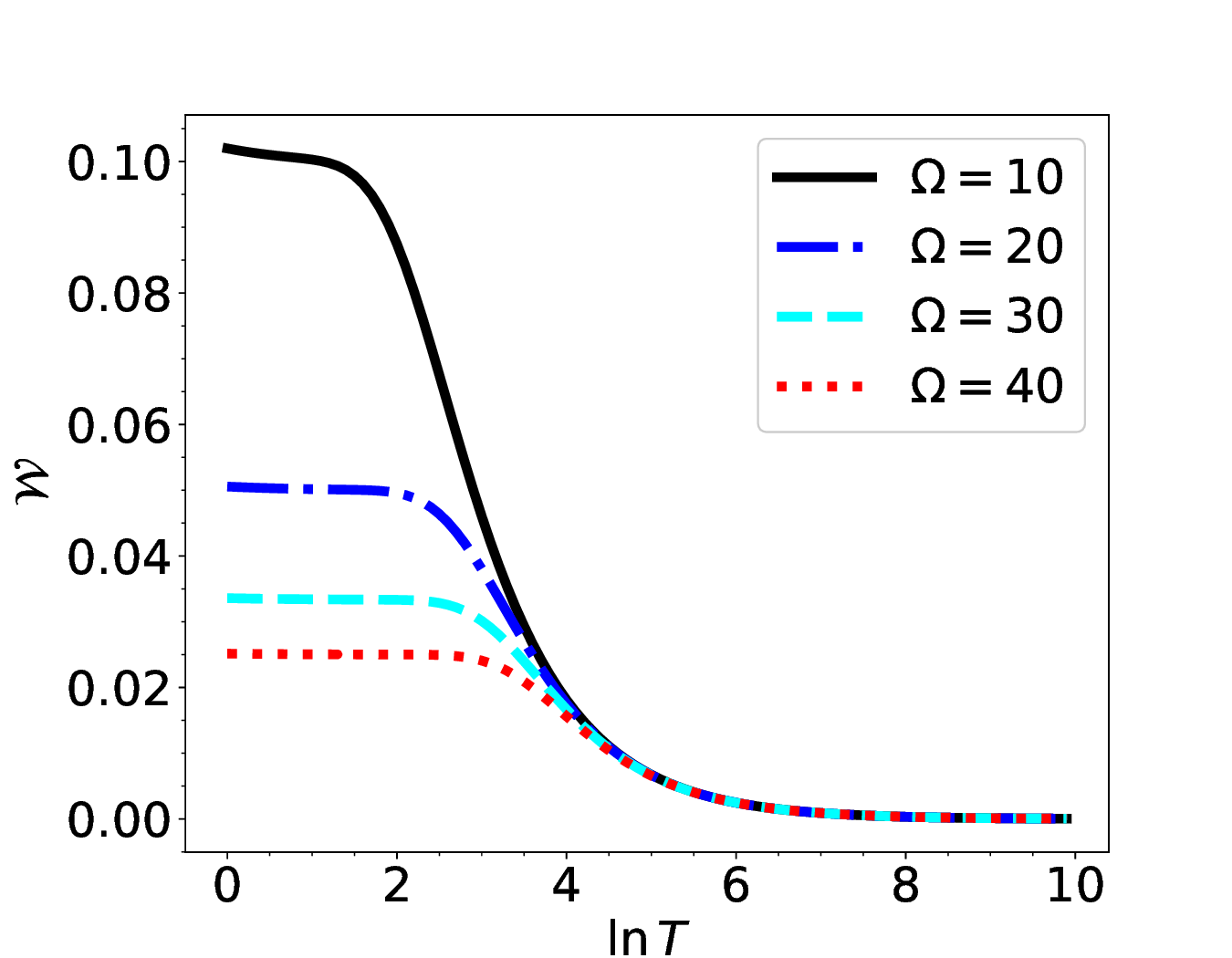}
    \caption{Ergotropy of Gravitational cat states in thermal bath as a function of temperature $T$ in the logarithmic scale for different values of gravitational interaction strength $\Omega$ with $\omega=1$. }
    \label{Fig2}
  \end{figure}
  In Fig.\ref{Fig2}, the ergotropy has been plotted as a function of temperature $T$ in the logarithmic scale for different values of gravitational interaction parameter $Omega$. It can be seen that the amount of work that can be extract from cat states in thermal bath decreases by increasing temperature. It can also be seen that the ergotropy decreases with increasing gravitational interaction strength. So, it can be concluded that temperature has negative role for extraction work from gravitational cat states in thermal bath.
  \begin{figure}[t]
\centering
    \includegraphics[width =0.8 \linewidth]{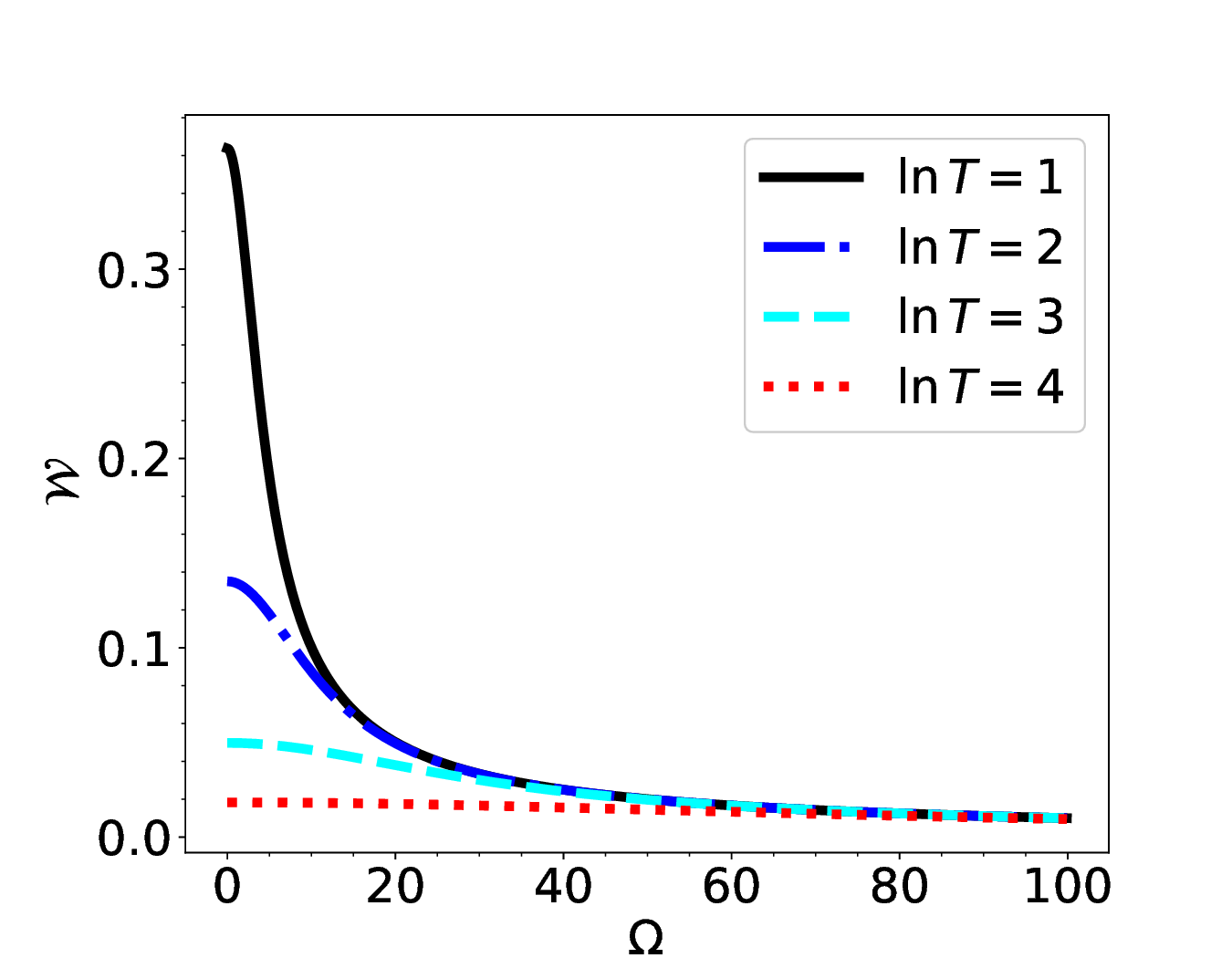}
    \caption{Ergotropy of gravitational cat states in thermal bath as a function of gravitational interaction strength $\Omega$  for different values of temperature in the logarithmic scale $\ln (T)$ with $\omega=1$.}
    \label{Fig3}
  \end{figure}
   In Fig. \ref{Fig3}, the ergotropy has been plotted as a function of gravitational interaction strength for different values of temperature in the logarithmic scale $\ln(T)$. As can be seen, the ergotropy decreases with increasing the interaction strength $\Omega$. In agree with Fig.\ref{Fig2}, It is observed that the ergotropy decreases with increasing temperature.  
   \begin{figure}[h]
\centering
    \includegraphics[width =0.8 \linewidth]{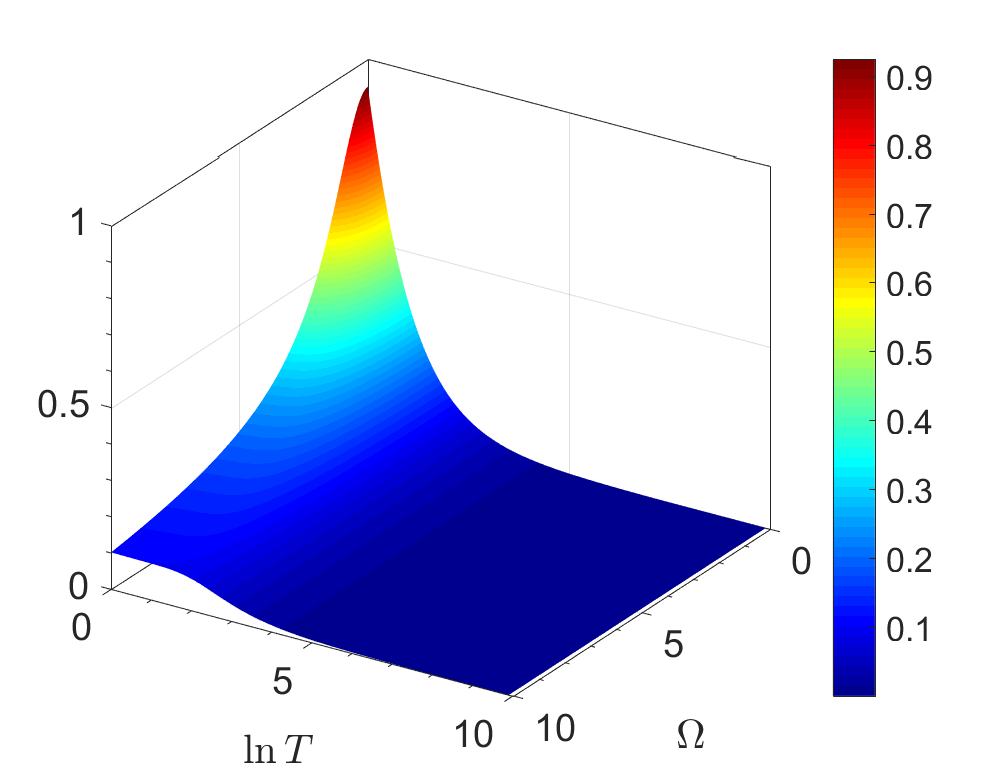}
    \caption{The ergotropy as a function of temperature in the logarithmic scale $\ln (T)$ and gravitational interaction strength $\Omega$ with $\omega=1$. }
    \label{Fig4}
  \end{figure}
Fig.\ref{Fig4}, shows the changes of ergotropy in terms of gravitational interaction strength and temperature. It is observed that the ergotropy decreases with increasing both temperature and gravitational interaction. 
\section{Conclusion}
In this work , we have studied the work that can be extracted  from quantum system of two massive particles confined in
two distinct double-well potentials in thermal bath. The state of the quantum system is known as gravitational cat state. It was observed that the temperature produced by thermal bath has negative role in work extraction procedure. In other words, it was showed that ergotropy as a measure of work extraction decreases with increasing temperature. It was also studied the effects of gravitational interaction strength on work extraction from gravitational cat state. It was observed that the amount of work that can be extracted from considered quantum system decreases with increasing gravitational interaction strength.





\end{document}